\documentclass{aa}
\usepackage{graphicx}
\usepackage{txfonts}
\begin{document}
   \title{Extreme gas properties in the most distant quasars}

   \author{R. Maiolino \inst{1}
          \and
	  E. Oliva \inst{1,2}
	  \and
	  F. Ghinassi \inst{2}
	  \and
	  M. Pedani \inst{2}
	  \and
	  F. Mannucci \inst{3}
	  \and
	  R. Mujica \inst{4}
	  \and
	  Y. Juarez \inst{4}
          }

   \offprints{R. Maiolino}

   \institute{
  	 INAF - Osservatorio Astrofisico di Arcetri,
   	Largo E. Fermi 5, I-50125 Firenze, Italy -
              \email{maiolino@arcetri.astro.it}
         \and
  	 INAF - Telescopio Nazionale Galileo,
   	Calle Alvarez de Abreu, 70,
	38700 Santa Cruz de La Palma, Spain 
	 \and
	 Istituto di Radioastronomia, sezione di Firenze
	  - Largo E. Fermi 5, I-50125 Firenze, Italy
	 \and
	 Instituto Nacional de Astrof\'{i}sica \'Optica y Electronica,
 Puebla, Luis Enrique Erro 1, Tonantzintla, Puebla 72840,  Mexico
             }

   \date{Received ; accepted }

   \abstract{
   We present near-IR, low resolution spectra of eight of
   the most distant quasars known, with redshifts in
   the range 4.9$<$z$<$6.4. Half of these quasars
   are characterized by deep, broad and blueshifted
   absorption features associated
   with both high and low ionization species (CIV, SiIV, AlIII, MgII),
   i.e. they belong to the class of Broad Absorption Line (BAL) quasars,
   which are associated with powerful outflows of dense gas.
   Although the sample is small, the large
   fraction of BAL quasars, the depth and ionization state of the
   absorption features suggest that these most distant quasars
   are surrounded by a much larger amount of dense gas than lower redshift
   (z$<$4) quasars. The possible interpretation in terms of extremely high
   accretion rates and the association with the early formation
   of quasars and of their host galaxies is discussed.
   The absorption properties of the dust, associated with the gas along the
   line of sight,
   appear different with respect to
   lower redshift quasars, possibly indicating different dust physics
   at these highest redshifts.

      \keywords{Galaxies: active -- Galaxies: evolution --
       Galaxies: high-redshift -- quasars: absorption lines --
       quasars: general}
   }

   \maketitle
%

\section{Introduction}

The exceptional luminosities of quasars allow us to study
the properties of gas and dust in their circumnuclear region
even in the most distant systems known.
The strong emission lines excited by the nuclear source have
been widely used to infer the metallicity and the physical/dynamical
state of the gas (e.g. Hamman \& Ferland \cite{hamann99},
Verner et al. \cite{verner03}).
These studies have been extended to the most distant quasars known
at z$>$5, approaching the epoch of re-ionization and of the
formation of the first generation of stars. The UV rest-frame
emission line properties of the highest redshift quasars have
indicated little or no evolution of the gas metallicity
(e.g. Dietrich et al. \cite{dietrich03},
Maiolino et al. \cite{maiolino03},
Pentericci et al. \cite{pentericci02}) up to z=6.4, suggesting
that the first epoch of star formation occurred at z$>$9,
in these objects.
Recent sub-mm and mm observations have allowed the detection of dust
emission in some quasars at z$>$5, indicating that large amounts
of dust were already formed at these early epochs
(Bertoldi et al. \cite{bertoldi03}, Priddey et al. \cite{priddey03}).

Within this context, Broad Absorption Line (BAL) quasars offer
additional tools to investigate the circumnuclear medium.
These objects, which account for about 15\% of the whole
quasar population at low and intermediate redshifts
(Reichard et al. \cite{reichard03b}, Hewett \& Foltz \cite{hewett03}),
are characterized by deep and broad 
absorption features associated with UV resonant lines. Generally
the most prominent absorption line is CIV$\lambda$1549.
A small fraction of BAL quasars (16\%, or $\sim$2\% of all quasars
at low and intermediate redshifts)
also show absorption by low ionization
species (MgII$\lambda$2798 and often AlIII]$\lambda$1857), these are
called LoBALs (in contrast to HiBALs, which only show the high
ionization absorption lines). An even rarer class are FeLoBALs,
which also show prominent FeII and FeIII absorption.
The basic interpretation is that
in BAL quasars we are seeing the active nucleus through
a dense outflowing gas.
BALs are characterized by redder continua
with respect to non-BALs, which is commonly interpreted as reddening
by dust associated with the outflowing gas
(Reichard et al. \cite{reichard03b}, Brotherton et al. \cite{brotherton01}).
The reddening
is generally fit best by a SMC-like extinction curve,
at variance with more absorbed AGNs which are characterized by a flatter
extinction curve (Gaskell et al. \cite{gaskell03},
Maiolino et al. \cite{maiolino01a}, Maiolino,
Marconi \& Oliva \cite{maiolinooliva01}). LoBALs are
generally characterized by higher reddening than HiBALs
(Reichard et al. \cite{reichard03b}, Becker et al. \cite{becker00}).

Various interpretations have been given for the BAL phenomenon.
One possibility is that all quasars are characterized by an
outflowing wind, but occurring only in certain preferential
directions; within this scenario BALs and non-BALs are the same
kind of quasars but observed along different viewing angles
(Weymann et al. \cite{weymann91}, Elvis \cite{elvis00},
Ogle et al. \cite{ogle99}, Schmidt \& Hines \cite{schmidt99}).
Alternatively, BALs and non-BALs could be two distinct populations
of quasars (Surdej \& Hutsemekers \cite{surdej87},
Boroson \& Meyers \cite{boroson92}), with and without outflows.
Becker et al. (\cite{becker00}) suggested that BALs may trace the early stages
of quasar activity, while King (\cite{king03})
and King \& Pounds (\cite{kingpounds03})
suggested that strong winds may be
related to episodes of enhanced accretion.

We have observed in the near-IR (UV rest-frame) a sample
of eight quasars in the redshift range
4.9$<$z$<$6.4 (among the highest redshift quasars currently
known).
The spectra were
obtained with an instrumental setup that can cover the full near-IR
spectrum from 0.8$\mu$m to 2.4$\mu$m in a single exposure (and with
very high throughput).
We find that four of these objects are BALs, and two (probably three) of them
are LoBALs. Although the statistics are very limited, this result
suggests that the fraction of BALs and LoBALs at z$>$5 may
be higher than at lower redshift. We discuss the extreme properties
of the gas and dust absorption in these objects, in comparison
with lower redshift quasar. We infer that gas and dust in
these most distant quasars are likely to be different 
with respect to the lower redshift cases, and this is probably
related to the early evolutionary stage of these quasars and
of their host galaxies.


\section{Observations and data reduction}

The observations were obtained with the
Near Infrared Camera Spectrograph (NICS) at the Italian Telescopio
Nazionale Galileo (TNG), a 3.56 m telescope. Among the various imaging and
spectroscopic observing modes (Baffa et al. \cite{baffa01}), NICS offers a
unique, high sensitivity, low resolution observing mode, which
uses an Amici prism as a dispersing element (Oliva et al. \cite{oliva03}).
In this
mode it is possible to obtain the spectrum from 0.8$\mu$m to 2.4$\mu$m
in a single exposure.
The throughput of the Amici prism is nearly two times higher
than other more commonly used dispersers.
The spectral resolution with a 0$''$.75 slit, as it
was in our case, is 75
(4000~km~s$^{-1}$) and nearly constant over the whole
wavelength range. This observing mode is appropriate
to study the near-infrared continuum of faint sources
and to detect broad ($\ge 4000$ km s$^{-1}$) emission 
and absorption lines in faint quasars.

Observations were performed in three observing runs: 2002 November 7-9;
2003 February 25-29; and 2003 May 23-26.
Several quasars were observed more than
once on different nights to check for any instrumental or observational
artifacts in the individual spectra. Wavelength calibration was performed
by using an argon lamp and the deep telluric absorption
features. The telluric absorption was then removed by dividing
the quasar spectrum by a reference star spectrum (FV-GV) observed
at similar airmass. The intrinsic features and slope of the
reference star were then removed by multiplying the corrected quasar
spectrum by
a spectrum of the same stellar type from the library by Pickles et al.
(\cite{pickles98}),
smoothed to our resolution. Absolute flux calibration was obtained
by using the photometry on the acquisition image or through photometry
reported in the literature.

These spectra were also discussed in Maiolino et al. (\cite{maiolino03})
in relation
to the FeII bump and MgII emission, with the goal of measuring the
iron abundance as a function of redshift. Here we focus on the
detection of the UV absorption lines tracing BALs and on the continuum
shape. For this reason we restrict this investigation to the
quasars in the Maiolino et al. (\cite{maiolino03}) sample with z$>$4.9,
for which
any putative CIV absorption is redshifted into the spectral range
where Amici+NICS allows a reliable detection (for all the quasar at
lower redshift the spectrum on the blue side of CIV is too noisy to
infer the presence of any absorption).
We include also the data
of SDSSJ1044-0125 obtained previously at the same instrument
and already published in Maiolino et al. (\cite{maiolino01b}).
The eight quasars selected
with this redshift constraint are all from the Sloan Digital Sky Survey (SDSS).

The quasars in our sample are listed in Tab.~\ref{tab1}, where their
full IAU designations are reported. However, in the following we will adopt
a shortened version of their names.

\section{BAL quasars}

In this sample we identify four high redshift BAL quasars,
which are discussed in detail in this section.

We determine the continuum
slope and the {\it balnicity index} of the CIV absorption (which is similar
to the Equivalent Width expressed in km/s, as defined in Weymann et al.
\cite{weymann91}) by following the method described
in Reichard et al. (\cite{reichard03a}).
More specifically,
the average SDSS non-BALs quasar spectrum, which has a continuum
slope $\alpha = -1.61$ ($\rm F_{\lambda}\propto \lambda ^{\alpha}$),
is taken as a template and modified to match the observed
spectrum. The template
is smoothed to our resolution. Then the continuum shape
is modified to match the line-free regions of the observed
spectrum either 
by changing the spectral index (by multiplying the spectrum by
$\lambda ^{{\alpha}/-1.61}$) or by absorbing the template
with an SMC extinction curve.
The CIV emission intensity of the template may also
be modified to match the observed intensity and to better calculate
the balnicity index; in practice in our cases this is a very
difficult task, since the CIV emission is so heavily absorbed that it
is difficult to estimate its intrinsic flux and profile (this issue
is discussed more in detail below).

   \begin{figure*}[!]
   \centering
   \includegraphics[width=8.5truecm]{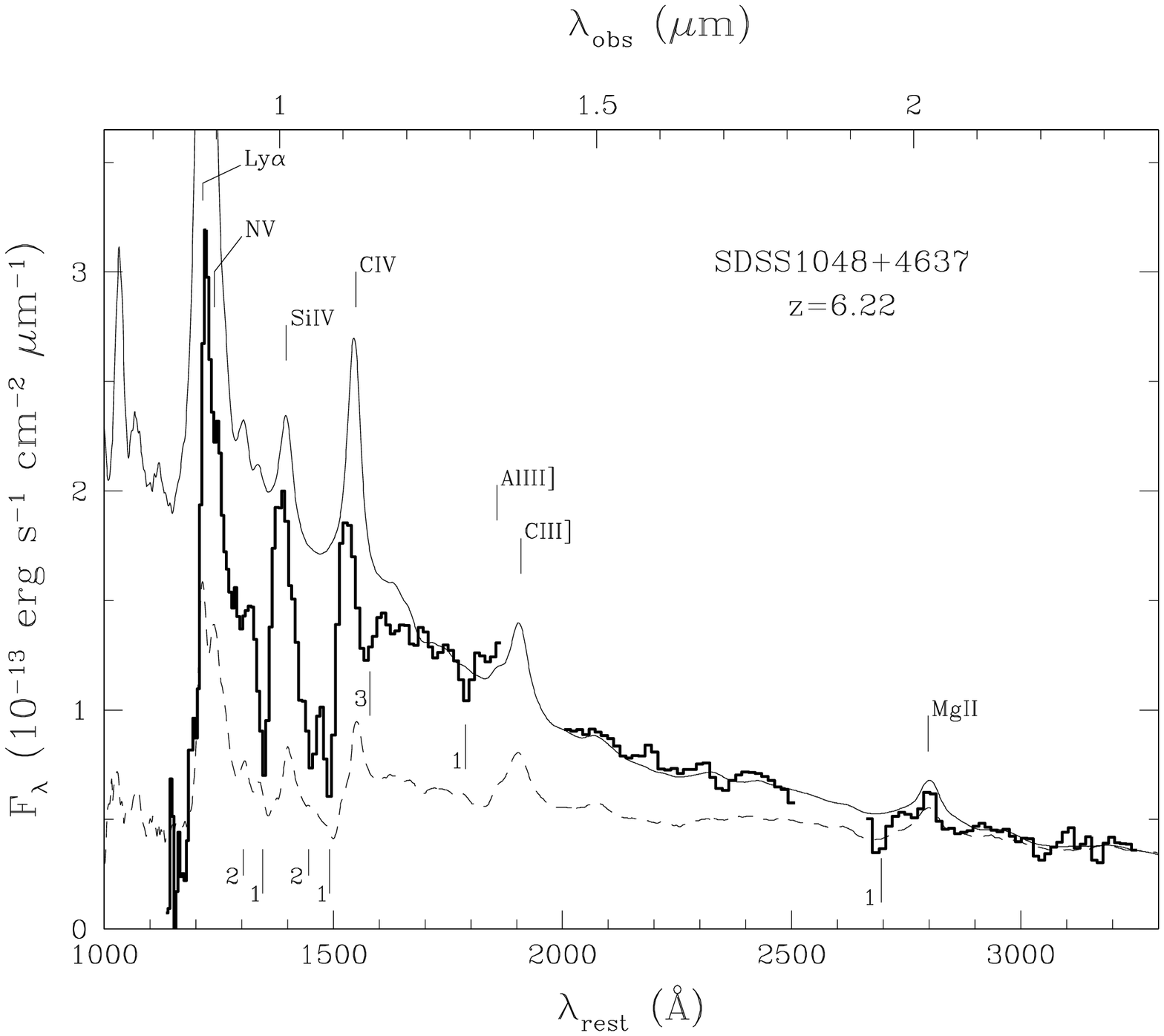}
   \includegraphics[width=8.5truecm]{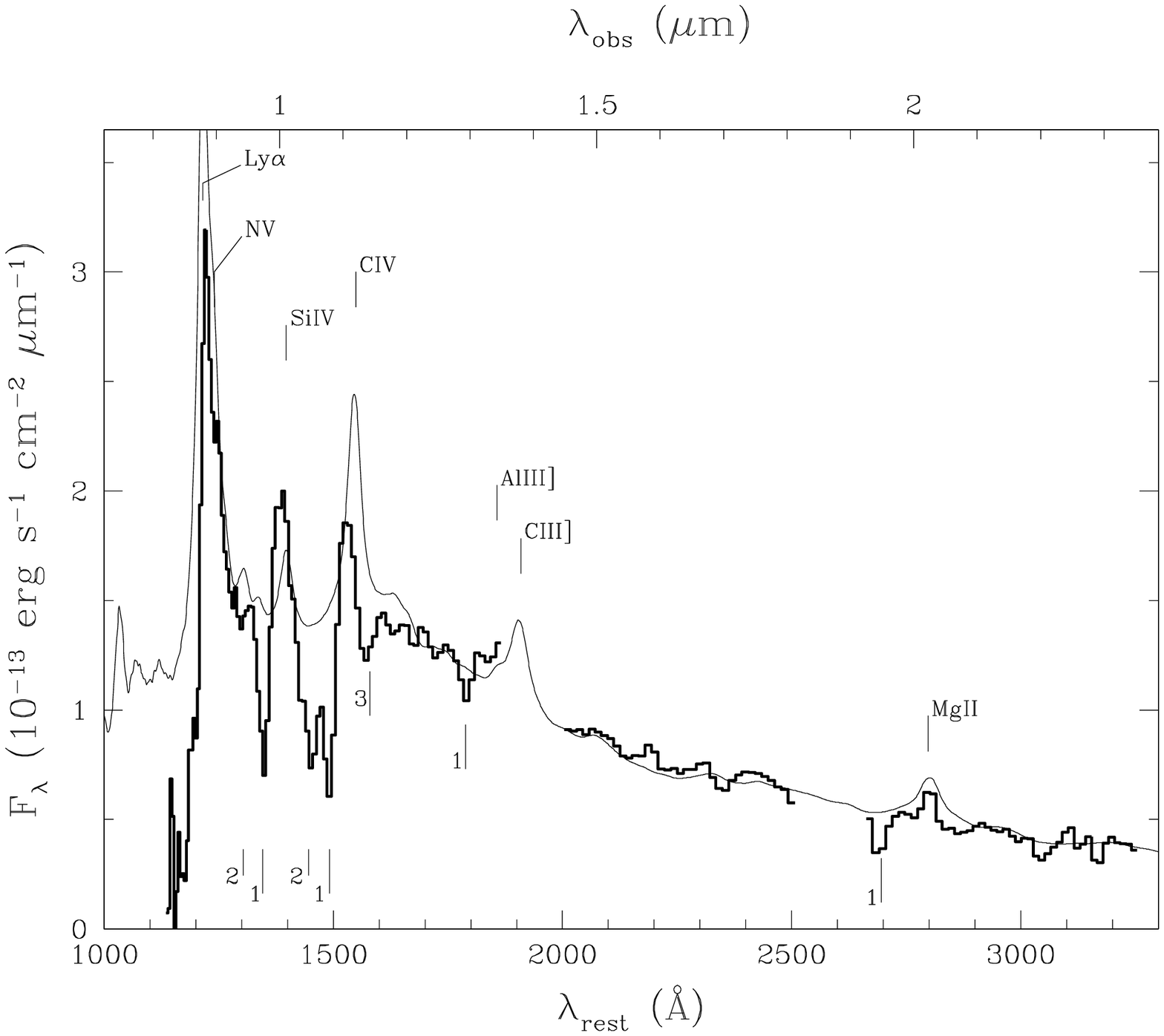}
      \caption{{\sl a)} Amici+Nics spectrum of SDSS1048+46, a LoBAL at
      z=6.22 (thick solid line). The thin line shows the non-BAL
      SDSS template adapted to a slope $\alpha =-2.1$.
      The thin dashed line shows the average spectrum of LoBAL
      quasars in the SDSS.
      The main emission lines are marked. The lower ticks mark
      the blueshifted absorptions at 11000~km/s
      and at 20000~km/s (systems {\it \#1} and {\it \#2} respectively).
      System {\it \#3} could be a
      redshifted absorption feature of CIV at $-$3800~km/s, alternatively
      it could be absorption by HeII$\lambda$1640 associated with
      system {\it \#1} (see text).
      {\sl b)} Same as {\sl a}, but where the template
      is absorbed by an extinction curve which is much steeper in the
      extreme UV (see text).
              }
         \label{fig1}
	 \begin{picture}(100,-100)
	 \put(40,245){\bf (a)}
	 \put(285,245){\bf (b)}
	 \end{picture}(100,-100)
   \end{figure*}

\subsection{SDSS1048+46: the most distant LoBAL quasar (z=6.22)}

At a redshift of 6.22 this
quasar is now the highest redshift BAL known.
Fan et al. (\cite{fan03}) suggested that, based on a hint of SiIV absorption
in the optical spectrum, this might be a BAL.
The Amici+NICS spectrum is shown in Fig.~\ref{fig1}a and clearly
displays a very deep and very broad CIV absorption. The
absorption profile is resolved and displays a clear double trough,
a feature which is commonly observed in LoBALs. The
two main troughs have a blueshift of 11000 and 20000 km/s, which are
labeled as systems {\it 1} and {\it 2}, respectively,
in Fig.~\ref{fig1}a.
At the lower velocity shift (11000 km/s) there is also a deep
SiIV absorption and absorption of AlIII] and MgII, which identify
this object as a LoBAL.

In Fig.~\ref{fig1}a the thin solid line shows the non-BAL template where the
slope has been adapted to match
the line-free continuum regions (redward of CIV). We have to force
a slope of $\alpha = -2.1$, bluer than the average SDSS spectrum
(but in agreement with the slope of dereddened BALs, as discussed below).
There is a strong mismatch between the
template and the observed spectrum blueward of CIV which might be
ascribed to the very deep absorption by the UV lines (but see below
for an alternative explanation).
The continuum slope redward of CIV would indicate
that this BAL is not subject to any dust reddening.
This result is at odds with what observed in lower redshift BALs,
and specifically
LoBALs, which are systematically reddened. To illustrate this difference the
thin dashed
line in Fig.~\ref{fig1}a shows the average LoBAL spectrum obtained by the SDSS
(Reichard et al. \cite{reichard03a}),
which is clearly much redder than the spectrum of
SDSS1048+46. {\it All} LoBALs quasars observed at lower redshift
have an observed spectral slope redder than observed in SDSS1148,
more specifically at z$<$4
{\it all} LoBAL quasars in the SDSS have an observed $\rm \alpha  > -1.9$.

Another possible explanation to fit the observed spectral shape
would be to assume that the flattening blueward
of 1600\AA \ is not due to broad line absorption but
to dust reddening. However, a SMC extinction curve
does not work, since it would make the whole spectrum redder. Even
steeper extinction curves, as those inferred in some rare BALs, would
be unable to reproduce the observed continuum shape. The only possibility
would be a (speculative)
extinction curve which is relatively flat at $\lambda > 1500$\AA \ 
and steeply rising at $\lambda < 1500$\AA \ (never observed so far).
We do not discuss in detail
this possibility, since it will be
investigated more in detail in a forthcoming paper (Maiolino et al., in prep),
but in Fig.~\ref{fig1}b we only show the effect of such an unusual extinction
curve on the continuum
shape. Obviously tackling this issue will require higher spectral resolution
over the whole band 0.8--1.3$\mu$m to disentangle the effects of broad
absorption lines from the true continuum curvature.

On the the very blue spectral shape redward of 1500\AA, with $\rm \alpha  =-2.1$
it is most interesting to note that the 'intrinsic' 
LoBALs spectral index inferred
by Reichard et al. (\cite{reichard03b}) is $-2.01$ 
(altough there is some degeneracy with reddening). Based also
on the analysis of their emission line properties, Reichard et al.
(\cite{reichard03b})
suggested that BALs may, in fact, be (intrinsically) bluer than the 
average quasar, despite being dust reddened. The lack of
dust absorption (at least reward of 1500\AA) in SDSS1048+46 allows us
to observe directly the intrinsic slope of this BAL and the resulting very blue
slope strongly supports the finding of Reichard et al. (\cite{reichard03b})
indirectly inferred for lower redshift BALs QSOs.

The balnicity index of CIV cannot be measured very accurately, because
it is not possible to properly recover the intrinsic profile of the CIV emission
prior of absorption. For this reason we derive
a balnicity index with a relatively large error, which is not due to
signal-to-noise but
dominated by the uncertainty on the CIV emission profile.
We obtain: BI(CIV)$=$6500$\pm$1100~km/s. As discussed in Sect.6,
this high value is located
on the high balnicity indices tail of the BALs observed
at lower redshift (Reichard et al. \cite{reichard03a}).

Another interesting feature of this BAL is the presence of
absorption even on the red side of CIV (system {\it \#3} in Fig.~\ref{fig1}).
If associated with redshifted CIV
the corresponding receding velocity would be about $-$3800~km/s,
such a redshifted trough was never observed at lower redshift
(only a weak ``droop'' was found on the red wing of CIV of
the average spectrum of BAL quasars
by Reichard et al. (\cite{reichard03b}), but
only after a careful subtraction of the emission line profile).
The redshifted trough may be explained in a scenario
of a rotating disk wind whose size (or at least part of it) is comparable
with the continuum emitting region. A similar scenario was invoked by
Hall et al. (\cite{hall02}) to explain the suppression of the red wing
of CIV in lower redshift BALs.
However, we note that the redshifted absorption may well be tracing
{\it inflowing} gas along the line of sight,
again under the assumption that the size of the
absorber is comparable with the emitting source. An alternative
explanation is that system {\it \#3} is actually absorption of
HeII${\lambda}$1640 associated with system {\it \#1}. Absorption
by HeII would require extreme gas properties in terms of opacity.
However, recently HeII blueshifted absorption has been detected in one
SDSS quasar (Hall et al. \cite{hall04}). Higher spectral resolution observations
of this object are certainly required to further investigate the
nature of this absorption system.

Finally we note that the SiIV and CIV emission lines appear significantly
blueshifted with respect to the template at z$=$6.22, which may suggest
a lower redshift for this source. However higher spectral resolution is
required to better disentangle absorption and emission components and
to better determine the redshift of this system. If CIV, in particular, is
really blueshifted with respect to the rest frame, this may be consistent
with the intrinsically bluer continuum of this QSO, in keeping with
the finding by Reichard et al. (\cite{reichard03b}) that bluer QSOs tend to be
associated with blueshifted CIV.

   \begin{figure}
   \centering
   \includegraphics[width=8.5truecm]{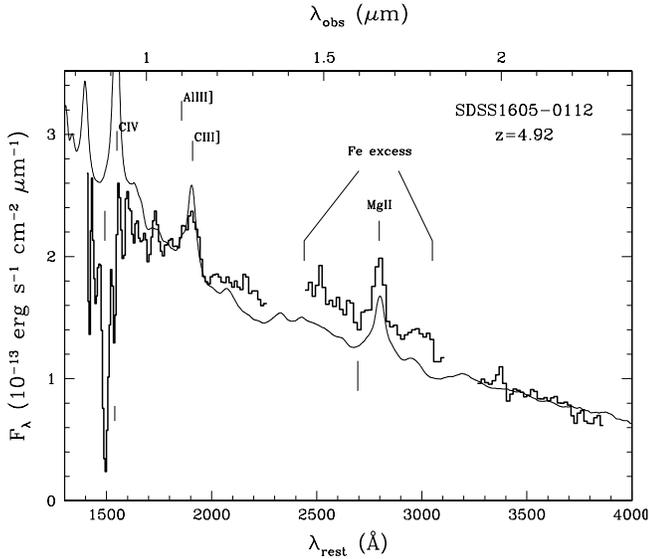}
      \caption{Amici+Nics spectrum of SDSS1605-01, a LoBAL at
      z=4.92 (thick solid line). The thin line shows the non-BAL
      SDSS template reddened with an SMC extinction curve with
      $\rm E_{B-V}=0.03$.
      The main emission lines are marked (though CIV is essentially
      totally ``eaten'' by the absorption).
      The long ticks mark
      the blueshifted absorptions at 11000~km/s for CIV (the deepest
      trough) and MgII. The shorter thick marks a shallower absorption
      trough blueshifted by 2000~km/s.
      Note an excess of FeII emission, not uncommon among the most
      extreme BALs.
              }
         \label{fig2}
   \end{figure}

\subsection{SDSS1605-01: an extreme LoBAL at z=4.92}

The Amici+NICS spectrum of this extreme BAL is shown in Fig.~\ref{fig2}.
This has the deepest CIV absorption of all BALs found
at high redshift. Its balnicity index of about 9300$\pm$2000~km/s
is among the highest value ever found among BALs known so far
(again, the large error on the balnicity index is dominated
by the large uncertainty on the CIV emission profile).
The CIV absorption is so deep that it has ``eaten''
nearly all of the CIV emission.
Balnicity indices higher than 8000~km/s
are only found among LoBAL quasars. Indeed, also in this object
there is indication of MgII absorption (though with a low S/N),
confirming the LoBAL nature of this object. 

This is the only BAL clearly characterized by a redder slope than
the average of quasars. The thin line in Fig.~\ref{fig2} shows the
non-BAL template reddened with an SMC extinction curve to match
the observed continuum slope. A reddening of $\rm E_{B-V}=0.03$
is required. Alternatively the spectrum can be fitted
by forcing the template to have a flatter power law and, more
specifically $\alpha =-1.35$. 

The fit in Fig.~\ref{fig2} suggests a strong excess of FeII emission. This excess
is unlikely to be a flux calibration problem, both because the
observing mode adopted by us prevents this kind of calibration problems
between the IR bands (since the full spectrum is obtained simultaneously),
and because this object was observed in two different nights and
both spectra show the same feature.
As noted by Weymann et al. (\cite{weymann91}), Fe emission
in excess of the average is commonly associated with BAL quasar
with high balnicity index.

Finally, we note that this object may also be characterized by redshifted
absorption, similar to the previous case. However, higher resolution
observations are required to clearly disentangle emission features from
possible redshifted absorption troughs.

   \begin{figure}
   \centering
   \includegraphics[width=8.5truecm]{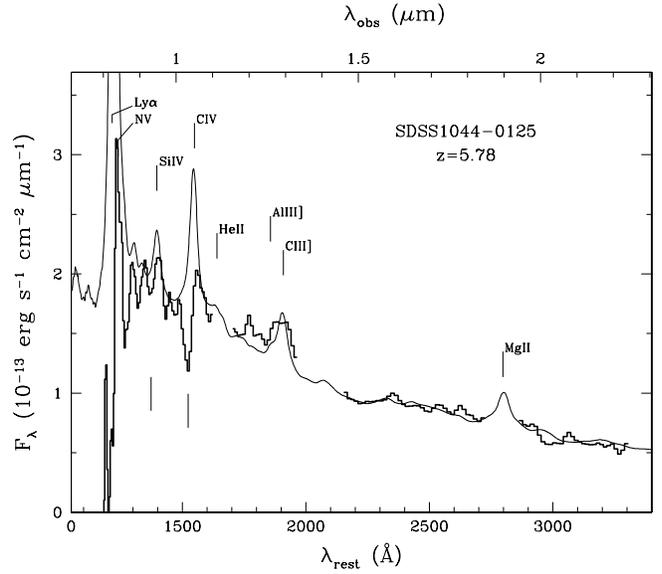}
      \caption{Amici+Nics spectrum of SDSS1044+01, a HiBAL at
      z=5.78 (thick solid line). The thin line shows the non-BAL
      SDSS template adapted.
      The main emission lines are marked. The lower ticks mark
      the blueshifted absorptions at 5000~km/s. 
              }
         \label{fig3}
   \end{figure}

\subsection{SDSS1044-01: the most distant HiBAL quasar (z=5.78)}

The Amici+NICS spectrum of this quasar was already shown in Maiolino
et al. (\cite{maiolino01b}).
We present again the spectrum here to extract the continuum
slope and balnicity index with the same template spectrum as the other
quasars. In Fig.~\ref{fig3} we show the observed spectrum along with the SDSS non-BAL
template (thin line).
Our spectrum is best fit by a redshift of 5.78 rather than
5.74 derived previously, but the difference my be caused by the
blending with the absorption features. The continuum slope appears only
slightly redder than the average of
the SDSS non-BAL quasars: $\alpha  = -1.55$ (though consistent
with the unreddened value of -1.62 within the uncertainties).  Note that in
Maiolino et al. (\cite{maiolino01b})
we found a better match (both for the continuum and
for the emission lines) with the
QSO template obtained by Francis et al. (\cite{francis91}).
However, even the slope of $-1.55$
is still significantly bluer than the average slope found in HiBAL quasars
at lower redshift, $\langle \alpha  (HiBAL, z<4)\rangle = -1.39$,
suggesting little or no dust reddening in this object.

The balnicity
index\footnote{Goodrich et al. (\cite{goodrich01})
obtain a much lower balnicity index
for this quasar because they integrated starting from
5000~km/s blueward of the peak, while the formal balnicity index has
to be calculated starting 3000~km/s blueward of the peak (Weymann et al.
\cite{weymann91}).}
is 1950$\pm$250~km/s. 

This BAL was observed with XMM by Brandt et al. (\cite{brandt01})
who derive a very steep optical--to--X-ray spectral index $\alpha _{OX}$,
which is rather typical of BAL quasars as a consequence of strong
gas absorption which suppresses the X-ray emission.

   \begin{figure}
   \centering
   \includegraphics[width=8.5truecm]{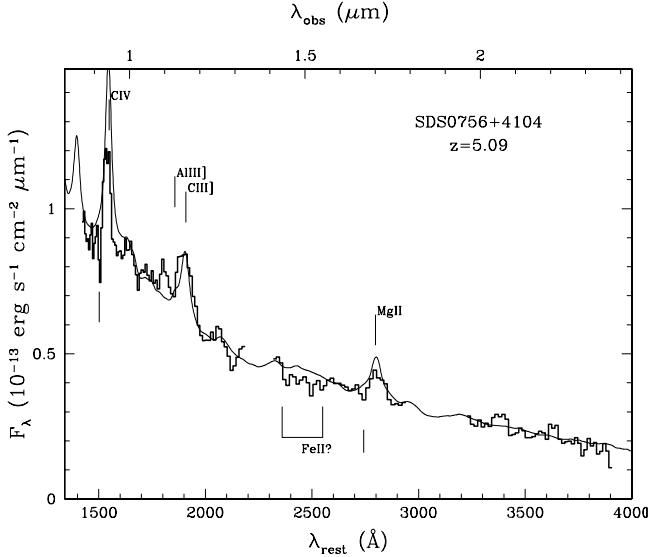}
      \caption{Amici+Nics spectrum of SDSS0756+41, a BAL at
      z=5.09 (thick solid line). The thin line shows the non-BAL
      SDSS template adapted.
      The main emission lines are marked (though CIV and CIII]
      show an apparent blueshift with respect to the template).
      The lower ticks mark 
      a CIV blueshifted absorption at 9000~km/s and a possible MgII
      absorption blueshifed by 6000~km/s. Possible absorption
      by FeII is also marked.
              }
         \label{fig4}
   \end{figure}

\subsection{SDSS0756+41 (z=5.09)}

The Amici+NICS spectrum of this quasar is shown in Fig.~\ref{fig4},
which is significantly noisier with respect to the previous spectra.
Our spectrum suggests a slightly lower redshift of 5.08 than
previously reported and both CIV and CIII] appear
blueshifted with respect to the template.
There is a blueshifted absorption (9000~km/s) of CIV 
with balnicity index of 270$\pm$60~km/s. This absorption is much
weaker than in the previous cases, but still quite significant (5$\sigma$),
also given
that the absorption is consistently observed in three different spectra
(taken on different nights), which were coadded to produce the final spectrum
shown in Fig.~\ref{fig4}. Some absorption may be observed blueward of MgII,
but with a slightly different velocity (6000~km/s). We also find
a dip in the blue region of the FeII bump, which might trace
Fe absorption and would identify this as
a FeLoBAL. However, higher S/N spectra are required to confirm
this finding.
FeLoBAL with weak or no CIV absorption have been found in the
SDSS (Reichard et al. \cite{reichard03a}).

The continuum slope is slightly bluer than the non-BAL
template ($\alpha =-1.67$), again indicating no dust
reddening.

There is an emission feature at 1800\AA , but we note that it
may be attributed to imperfect subtraction of a bright sky
emission feature in this spectral region.

   \begin{figure*}
   \centering
   \includegraphics[width=14truecm, angle=-90]{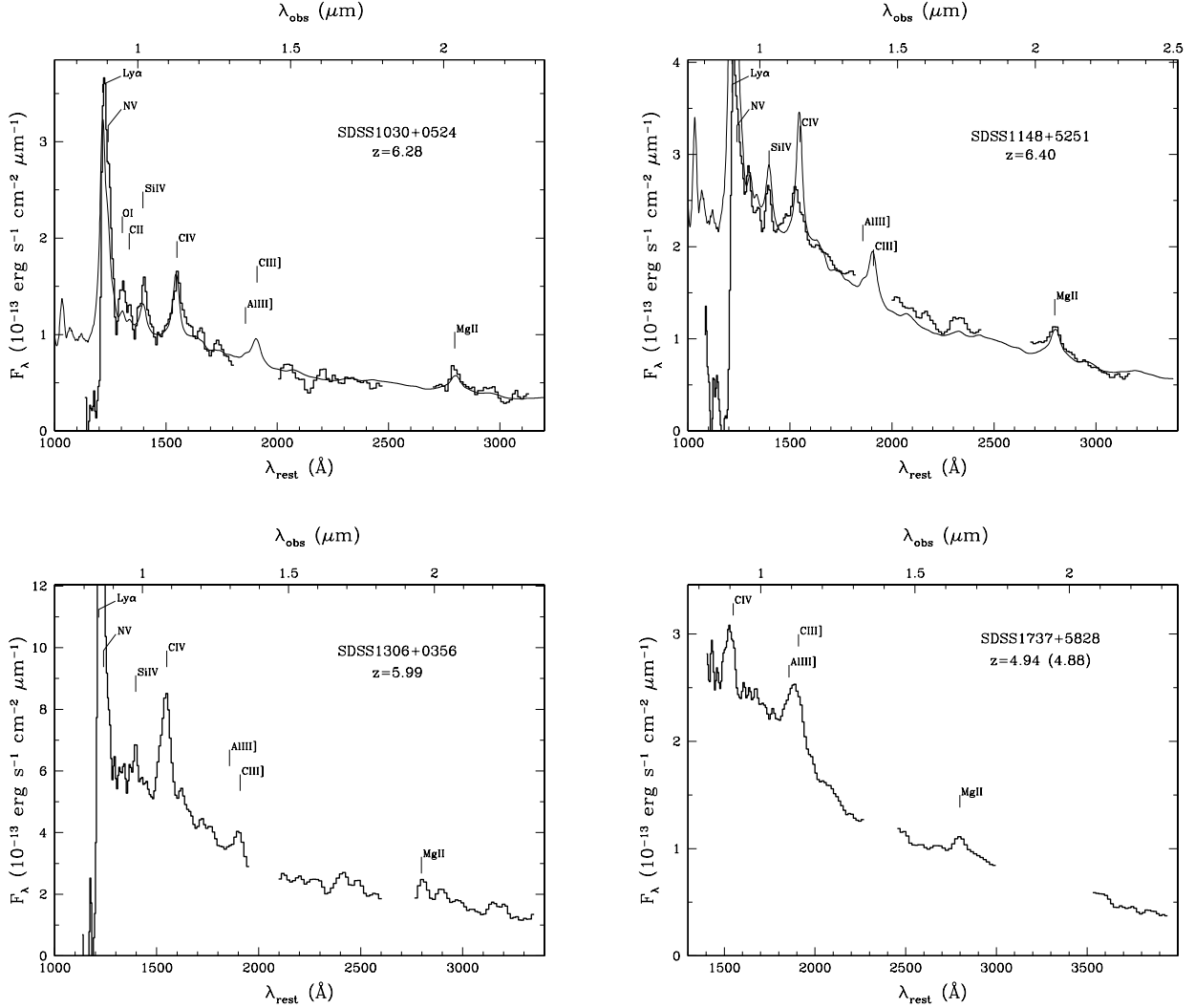}
      \caption{Amici+Nics spectra of non-BAL quasars at
      high redshift.
              }
         \label{fig5}
   \end{figure*}

\begin{table*}[!]
\caption{Summary of the properties derived from the spectra
of our sample}
\begin{tabular}{lllllll}
\hline
Name & z & type & BI(CIV)$^a$ & $\alpha $$^b$ & E(B-V)$^c$ & $\rm M_{i^*}$$^d$\\
\hline
SDSSJ075618.14+410408.6 & 5.08$^e$ & BAL$^e$ & 270$\pm$60$^g$    & $-$1.67 &  -- & --26.6 \\
SDSSJ103027.10+052455.0 & 6.28 &       &               & $-$1.55 & 0.001&  --28.1 \\
SDSSJ104433.04-012502.2 & 5.78$^h$ & HiBAL & 1950$\pm$250$^g$  & $-$1.55 & 0.001 & --28.4\\
SDSSJ104845.05+463718.3 & 6.22$^i$ & LoBAL & 6500$\pm$1100$^g$ & $-$2.10  &  --$^j$ & --27.4 \\
SDSSJ114816.64+525150.3 & 6.40 &       &               & $-$1.70 &  --  &  --28.5\\
SDSSJ130608.26+035626.3 & 5.99 &       &               & $-$1.80  &  --  &  --27.8\\
SDSSJ160501.21-011220.6 & 4.92 & LoBAL & 9300$\pm$2000$^f$ & $-$1.35 & 0.03 & --27.7\\
SDSSJ173744.88+582829.6 & 4.88$^k$ &       &               & $-$2.10 &  -- &  --26.9\\
\hline
\end{tabular}
\vspace{0.1cm}\\
Notes:\\
$^a$ Balnicity index of the CIV absorption as defined in
   Weymann et al. (\cite{weymann91}), which is ``similar''
   to the Equivalent Width
   of the absorption line measured in km/s.\\
$^b$ Slope of the continuum ($\rm F_{\lambda}\propto \lambda ^{\alpha }$);
 this is calculated by comparison with the non-BAL SDSS template which
 has $\alpha  = -1.61$ (see text).\\
$^c$ Reddening to apply to the SDSS non-BAL template to match the
 observed spectrum; a SMC extinction curve is assumed (see text).\\
$^d$ Absolute magnitude in the rest frame i$^*$ band, obtained by using
the observed spectral slope and assuming
$\rm H_0 = 50~km~s^{-1}~Mpc^{-1}$,
$\rm \Omega _m=1$ and $\rm \Omega _{\Lambda}=0$ (to allow for a direct
comparison with Reichard et al. \cite{reichard03a}); it is not corrected
for intrinsic absorption of the QSOs.\\
$^e$ This is the redshift obtained by our spectrum, while the
redshift reported in the literature is 5.09.\\
$^f$ Possible FeLoBAL (see text).\\
$^g$ All errors on the balnicity indices are dominated by the uncertainty
 on the profile of the CIV emission.\\
$^h$ This is the redshift obtained by our spectrum, 
while Goodrich et al. (\cite{goodrich01}) report a redshift of 5.74.\\
$^i$  This is our best estimate of the redshift based on the MgII (which
is slightly different than the redshiftof 6.23 published by
Fan et al. \cite{fan03}) ; however,
the CIV and SiIV emission features may suggest an even lower redshift.\\
$^j$ The extreme UV continuum of this LoBAL may be fitted with dust
 reddening, but with an extinction curve totally different with respect
 to the SMC one adopted for all other BAL (see text for details).\\
$^k$ This is the redshift obtained by our spectrum,
  while the redshift reported in the literature is significantly
  different: z=4.94.

\label{tab1}
\end{table*}

\section{Non-BAL quasars at 4.9$<$z$<$6.4}

Four out of the eight quasar at z$>$4.9 do not show
indication of BAL-like absorption features. Their Amici+NICS
spectra are shown in Fig.~\ref{fig5}. However the limited
spectral resolution (which might cause a blending with the
emission lines) and the limited signal-to-noise do not allow us
to exclude the presence of absorption features with a
balnicity index below $\sim$200~km/s.
As listed in Tab.~\ref{tab1} the continuum of these non-BAL spectra
does not show indication of significant
reddening and, actually, they are generally bluer than
the average of non-BAL quasars found at lower redshift.
This finding suggests that, at least in
this small sample, there are no representatives of the
reddened quasar found at lower redshift (Richards et al. \cite{richards03}).

\section{The fraction of BAL quasars at 4.9$<$z$<$6.4}

As mentioned in the Introduction, the fraction of BAL quasars
at low redshift is about 15\%. Only 2\% of the quasar population
are LoBAL, while FeLoBAL quasars are even more rare.
Reichard et al. (\cite{reichard03b}) obtained that the fraction of BAL quasars
is constant up to z$<$4.

The very limited statistics of our sample does not allow us to draw
statistically firm conclusions on the fraction of BAL quasars
at high redshift. However, it is impressive that out of 8 quasars
4 are BAL. The probability that the fraction of BAL quasars in this small
sample is consistent with the fraction of 0.15 found at lower redshift
is only 2\%. Even more astonishing is that two (may be three)
of these objects are
LoBAL, which is an extremely rare class at low redshift. Note that
there are no obvious effects, in terms of spectral shape,
that would favor the selection of BAL quasars
at z$>$5 in the the SDSS survey (actually any dust reddening
would bias against BAL quasars).
These findings suggest that at z$\sim$5--6 the fraction of
BAL, and specifically the fraction of LoBAL, increases
dramatically. This might be associated with larger amount
of (accreting) gas surrounding the circumnuclear
region of these high redshift quasars, as discussed in sect.8.

Another possibility is that the larger fraction of BALs among
high redshift quasars is a consequence of a selection effect.
More specifically, high redshift QSOs are more luminous and, therefore,
probably accreting at higher rate. If BAL QSOs are associated with
high accretion rates,
as suggested by various studies (sect.1 and sect.8), an
implication would be that they are more likely to be selected
at higher redshift. Actually, at z$<$4 where larger samples
of quasars and BALs are available, there is no evidence that
the BAL fraction change with luminosity, although dust absorption
and variations of the spectral slope (Reichard et al. \cite{reichard03b})
may prevent the detection of such a dependence.
However, we have investigated
this possibility by comparing the luminosity
of our BAL quasars at z$\sim$5--6
with the luminosity of the BALs at lower redshift in Reichard et al. (\cite{reichard03b}).
We derived the absolute magnitude $\rm M_{i^*}$ in the rest frame $\rm i^*$ band
(which is tabulated also for the SDSS BAL QSOs at z$<$4), by using
the extrapolation of the slope measured in our spectra to derive
the K-correction. The resulting absolute magnitudes\footnote{For consistency
with Reichard et al. (\cite{reichard03b}) we adopt $\rm H_0 = 50~km~s^{-1}~Mpc^{-1}$,
$\rm \Omega _m=1$ and $\rm \Omega _{\Lambda}=0$.}
are reported in Tab.~\ref{tab1}.
If in Reichard et al. (\cite{reichard03b}) we select those
BALs at 2.5$<$z$<$4, a range where the statistics is still very high
and where the fraction of BAL is still estimated to be $\sim$15\%,
we obtain absolute magnitudes ranging from -26 to -28.8
with an average of
$\rm \langle M_{i^*}\rangle _{2.5<z<4}=-27.4$, and a 
standard deviation $\rm \sigma _{M_{i^*}}=0.66$. The comparison
with the absolute magnitudes in Tab.~\ref{tab1} indicates no obvious
differences in the luminosity distribution between the sample
of BALs at $\rm 4.9<z<6.4$ and those at $\rm 2.5<z<4$. Tab.~\ref{tab1} also shows
that the non-BAL quasars in our sample are even more luminous than BALs.
If one takes into account that BALs at lower redshift tend
to be absorbed by dust (which seems not to be the
case in our redshift sample, as discussed in sect.7) a correction
for dust absorption would probably make the low redshift sample
even more luminous. Summarizing, a contribution from a luminosity
effect on the larger fraction of BALs at high redshift cannot
be excluded, but probably it does not play a dominant role.

\section{The balnicity index at 4.9$<$z$<$6.4}

With four BAL quasars only it is not possible to derive
a distribution for the CIV balnicity index among quasars at z$>$4.9.
Nonetheless, it is most interesting to note that three of these
BALs have very high balnicity indices, which are
in the high tail of the distribution found for quasars at z$<$4 (Fig.7
of Reichard et al. \cite{reichard03a}). In particular two of the BAL in our sample
have a balnicity index higher than 6000~km/s, while at lower redshift (z$<$4)
only 2\% of BAL have balnicity index higher than this value.
This result is also indicative of larger amount
of circumnuclear gas in these high redshift quasars.

\section{Dust at 4.9$<$z$<$6.4}

One of the most interesting peculiarities of the BAL quasars found
at z$>$4.9 is that most of them appear not to be reddened by dust.
In particular most of them are bluer than the average slope of
the corresponding sub-class at z$<$4 (we recall that at low redshift
$\langle \alpha  [NonBAL]\rangle = -1.62$,
$\langle \alpha  [HiBAL]\rangle = -1.39$,
$\langle \alpha  [LoBAL]\rangle = -0.93$, Reichard et al. \cite{reichard03b}).
There are some deviations from this trend, in particular
Stern et al. (\cite{stern03}) have found evidence for significant reddening
in one quasar at z=5.8 and also, within our sample, one of the
two LoBAL quasars appears to be significantly reddened (although still
bluer than the average of low redshift LoBAL). However, the general
trend is to have bluer continuum slopes.
The most intriguing case is the most distant LoBAL (SDSS1048+46
at z=6.22, Fig.~\ref{fig1}) which is much bluer than any LoBAL found at
z$<$4.

One naive interpretation would be that at these high redshift we
are approaching the epoch when dust had no time to form in large
quantities, both because of the lower metallicities and because
AGB stars did not have enough time to inject dust into the ISM.
Therefore, these high redshift quasars could be relatively poor in dust.
On the contrary, powerful dust emission in the mm/submm
was detected in several of these quasars, and in particular
in three of the BAL quasars, indicating that these systems are
characterized by very large dust
masses (Priddey et al. \cite{priddey03}, Bertoldi et al. \cite{bertoldi03}).

A possible explanation of this puzzling result may be found
in the origin and evolution of dust in these very high redshift quasars.
Possibly, the gas density in the circumnuclear region of these primordial
quasars is much higher than at lower redshift. Higher gas density may
favor the growth of grains and yielding to a flatter extinction curve, which
would significantly reduce the reddening by dust. 
An additional or alternative
possibility is that at these extreme redshifts the main dust production
mechanism is different.
While at low redshift dust is predominantly produced in
AGB stars, at very high
redshift dust may be predominantly produced by SNe
(Schneider et al. \cite{schneider03},
Nozawa et al. \cite{nozawa03}, Todini \& Ferrara \cite{todini01}) or
even directly produced in the the BLR of quasars (Elvis et al. \cite{elvis02}):
the different dust production mechanism may
yield an extinction curve quite different than observed locally.
This issue will be discussed more in detail in a forthcoming paper
(Maiolino et al. in prep.).

\section{Discussion}

Although the small number of objects does not allow us to draw
strong conclusions, our results suggest that the physics of the
circumnuclear medium of the most distant quasars currently known,
at 4.9$<$z$<$6.4, is probably different than in lower redshift
quasars. The large fraction of BAL quasars suggest that either
the outflowing wind occurs over a much wider solid angle (in a scenario
where BAL and nonBAL quasars are unified through the viewing angle)
or that the fraction of quasars experiencing strong outflowing winds
is much larger at high redshift (in a scenario where the wind of BAL
quasars covers isotropically the continuum source). Both scenarios would
imply an evolution of quasars' circumnuclear medium at high redshift.
Both the extreme
depth of the absorption features and the identification of low ionization
species (MgII, AlIII]) suggest that the column density of
the outflowing gas is higher in these high redshift quasars.
Summarizing, quasars at 4.9$<$z$<$6.4 are probably expelling larger
amount of gas, through winds, than quasars at z$<$4.

As suggested by some authors (e.g. King \cite{king03},
King \& Pounds \cite{kingpounds03},
Mathur \cite{mathur01}) the strong winds observed in some AGN are probably
associated with large accretion rates. In particular, quasars characterized
by the strongest outflows (in terms of mass outflow rate)
may be accreting at or above the Eddington
rate. Within this context, the differences found by us between low and
high redshift quasars may reflect different evolutionary stages, where
the most distant quasars are accreting at a significantly higher rate.
Such an evolutionary scenario for these highest redshift BAL quasars
would also be consistent with the study of Becker et al. (\cite{becker00})
on radio loud quasar, who find that the BAL
phenomenon is preferentially associated with younger radio sources and,
therefore, tracing the early stages of quasar activity.
Recently Willott et al. (\cite{willott03}) and Lewis et al. (\cite{lewis03})
suggested
that the evolutionary scenario for BALs might be wrong based on the
lack of difference in terms of sub-mm emission between non-BALs and BALs.
However, they only probed the redshift range $\rm 1.0<z<2.6$. As
mentioned by Willott et al. (\cite{willott03}),
their finding may still fit within the 
evolutionary scenario if the sub-mm emission by the host galaxy
is not related to the BAL activity of the nucleus, at least in the
$\rm 1.0<z<2.6$ redshift range.

At z$\sim$6 we are probably probing the epoch of formation for the
first galaxies, and approaching the re-ionization epoch,
as inferred by the evolved metallicity of these systems (e.g.
Maiolino et al. \cite{maiolino03},
Dietrich et al. \cite{dietrich03},
Pentericci et al. \cite{pentericci02}).
Our results would support the scenarios of a co-evolution
between QSOs and host galaxies, which predict that the rapidly
forming spheroids at high redshift
would be associated with a large accretion rate
onto the central Black Hole as a consequence of the enhanced viscous
drag (e.g. Granato et al. \cite{granato04}, Umemura et al. \cite{umemura97}). The same models expect a strong
feedback by the quasars themselves in the form of strong outflows,
as actually observed by us. Although the qualitative agreement between
observations and models is remarkable, much effort is still required
for an accurate quantitative comparison.

\section{Conclusions}

We have presented near-IR spectra covering the whole 0.8--2.4$\mu$m range
for eight among the most distant quasars known,
and specifically in the redshift range 4.9$<$z$<$6.4. At these redshifts
our spectra sample the UV rest-frame spectral region and, in particular,
include several prominent UV lines such as CIV, AlIII], CIII], MgII and
the UV FeII bump. The observational results strongly suggest that the
physical properties of the
circumnuclear medium of these most distant quasars are different
with respect to lower redshift quasars. More specifically,
the observational results can be summarized as follows:
\begin{enumerate}

\item Half of the quasars show broad and deep (blueshifed) absorption
 of CIV, i.e. belong to the class Broad Absorption Line (BAL) quasars,
 which are characterized by large quantities of
 gas outflowing at high velocities.
 At lower redshift (z$<$4) the fraction of BAL quasars is about 15\%.
 Our finding strongly suggests that the fraction of BAL quasars increases
 at z$>$4.9. We cannot exclude a possible luminosity dependence
 of the BAL fraction, although probably it does not play a
 major role.

\item Two (possibly three) of the BAL quasars also show absorption
 associated
 with low ionization lines (MgII and AlIII]), i.e. belong to the subclass of
 ``Low Ionization BAL'' (LoBAL). This class of quasars is very rare
 at lower redshifts, accounting only for 2\% of the whole quasar population
 at z$<$4. Our finding suggests that LoBAL may be much more common at
 z$>$4.9.

\item The CIV absorption features are on average deeper than found
in lower redshift BAL quasars. This finding, along with the presence
of low ionization species in absorption, indicates that the column
density of the outflowing gas is larger than in lower redshift quasars. 

\item The most distant BAL quasar (SDSS1048+46, a LoBAL at z=6.22) is
 also characterized by an absorption feature on the
 red side of CIV. This might be absorption due to redshifted
 CIV,
 or blueshifted absorption associated with HeII$\lambda$1640
 (observed only in one case at lower redshift).

\item The continuum of the quasars in our sample is on average
bluer than the quasars at lower redshift. The most interesting case
is the most distant LoBAL quasar which is bluer than any quasar of
the same class found at lower redshift (although an intriguing
bending of the spectrum
at $\lambda < 1500$ is observed) . These findings suggest little,
or very low, dust reddening with
respect to what observed in lower redshift quasars. Yet,
dust is present in large quantities, as inferred by the mm/submm detections.

\end{enumerate}

These observational results suggest the following conclusions
about the physics of the most distant quasars:\\

i) The large fraction of BAL quasars, the depth and ionization
state of the absorption systems suggest that, at z$\sim$5--6,
quasars are surrounded by larger amount of denser gas with respect
to lower redshift quasars.\\

ii) The presence of strong outflows is likely associated with high
accretion rates in these primordial quasars. Strong quasars winds
and high accretion rates are expected by the models of co-evolution
of quasars and host galaxies at high redshifts.\\

iii) The dust extinction curve in these distant quasars is probably
different than lower redshift quasars. This may reflect different
evolution and formation mechanisms of dust grains at z$>$5.
This issue will be discussed more in detail in a forthcoming
paper (Maiolino et al., in prep).

\begin{acknowledgements}
 We thank G. Richards and P. Hall for very useful comments.
      This work was partially supported by the Italian
      Ministry of Research (MIUR) 
      and by the National Institute for Astrophysics (INAF).
\end{acknowledgements}


\begin{thebibliography}{}

\bibitem[2000]{becker00} Becker, R.~H., White, 
R.~L., Gregg, M.~D., Brotherton, M.~S., Laurent-Muehleisen, S.~A., \& Arav, 
N.\ 2000, \apj, 538, 72 


\bibitem[2001]{baffa01} 
Baffa, C.~et al.\ 2001, \aap, 378, 722 

\bibitem[2003]{bertoldi03} Bertoldi, F., Carilli, 
C.~L., Cox, P., Fan, X., Strauss, M.~A., Beelen, A., Omont, A., \& Zylka, 
R.\ 2003, \aap, 406, L55

\bibitem[1992]{boroson92} Boroson, T.~A.~\& 
Meyers, K.~A.\ 1992, \apj, 397, 442

\bibitem[2001]{brandt01} Brandt, W.~N., 
Guainazzi, M., Kaspi, S., Fan, X., Schneider, D.~P., Strauss, M.~A., 
Clavel, J., \& Gunn, J.~E.\ 2001, \aj, 121, 591

\bibitem[2001]{brotherton01} Brotherton, M.~S., 
Tran, H.~D., Becker, R.~H., Gregg, M.~D., Laurent-Muehleisen, S.~A., \& 
White, R.~L.\ 2001, \apj, 546, 775

\bibitem[2003]{dietrich03} Dietrich, M., Hamann, F., 
Appenzeller, I., \& Vestergaard, M.\ 2003, \apj, 596, 817

\bibitem[2000]{elvis00} Elvis, M.\ 2000, \apj, 545, 63

\bibitem[2002]{elvis02} Elvis, M., 
Marengo, M., \& Karovska, M.\ 2002, \apjl, 567, L107

\bibitem[2003]{fan03} Fan, X.~et al.\ 2003, \aj, 
125, 1649

\bibitem[1991]{francis91} Francis, P.~J., at al. \ 
1991, \apj, 373, 465

\bibitem[2003]{gaskell03} Gaskell, C.~M., et al.
 \ 2003, ApJ, submitted (astro-ph/0309595)

\bibitem[2001]{goodrich01} Goodrich, R.~W.~et 
al.\ 2001, \apjl, 561, L23

\bibitem[2004]{granato04} Granato, G.~L., Silva, L.,
De Zotti, G., Bressan, A., Danese, L.\ 2003, \apj, 600, 58

\bibitem[2002]{hall02} Hall, P.~B.~et al.\ 2002, 
\apjs, 141, 267 

\bibitem[2004]{hall04} Hall, P.~B.~et al.\ 2004, 
in Multiwavelength AGN Surveys, eds. R. Mujica and R. Maiolino,
World Scient. Publ., in press (astro-ph/0403347)

\bibitem[1999]{hamann99} Hamann, F.~\& 
Ferland, G.\ 1999, \araa, 37, 487 

\bibitem[2003]{hewett03} Hewett, P.~C.~\& 
Foltz, C.~B.\ 2003, \aj, 125, 1784

\bibitem[2003]{king03} King, A.\ 2003, \apjl, 596, L27

\bibitem[2003]{kingpounds03} King, A.~R.~\& Pounds, 
K.~A.\ 2003, \mnras, 345, 657

\bibitem[2003]{lewis03} Lewis, 
G.~F., Chapman, S.~C., \& Kuncic, Z.\ 2003, \apjl, 596, L35

\bibitem[2001a]{maiolino01a} Maiolino, R., Marconi, 
A., Salvati, M., Risaliti, G., Severgnini, P., Oliva, E., La Franca, F., \& 
Vanzi, L.\ 2001a, \aap, 365, 28

\bibitem[2001]{maiolinooliva01} Maiolino, 
R., Marconi, A., \& Oliva, E.\ 2001, \aap, 365, 37

\bibitem[2001b]{maiolino01b}
Maiolino, R., Mannucci, F., Baffa, C., Gennari, S., \& Oliva, E.\ 
2001b, \aap, 372, L5 

\bibitem[2003]{maiolino03} Maiolino, R., Juarez, 
Y., Mujica, R., Nagar, N.~M., \& Oliva, E.\ 2003, \apjl, 596, L155

\bibitem[2001]{mathur01} Mathur, S.\ 2001, \aj, 122, 
1688

\bibitem[2003]{nozawa03} Nozawa, T., et al.\ 2003, ApJ, in press
(astro-ph/0307108)

\bibitem[1999]{ogle99} Ogle, P.~M., Cohen, M.~H., 
Miller, J.~S., Tran, H.~D., Goodrich, R.~W., \& Martel, A.~R.\ 1999, \apjs, 
125, 1 

\bibitem[2003]{oliva03} Oliva, E.\ 2003, Memorie della 
Societa Astronomica Italiana, 74, 118

\bibitem[2002]{pentericci02} Pentericci, L.~et 
al.\ 2002, \aj, 123, 2151

\bibitem[2003]{priddey03} Priddey, R.~S., Isaak, 
K.~G., McMahon, R.~G., Robson, E.~I., \& Pearson, C.~P.\ 2003, \mnras, 344, 
L74

\bibitem[1998]{pickles98} 
Pickles, A.~J.\ 1998, \pasp, 110, 863 

\bibitem[2003a]{reichard03a} Reichard, T.~A.~et 
al.\ 2003a, \aj, 125, 1711

\bibitem[2003b]{reichard03b} Reichard, T.~A.~et 
al.\ 2003b, \aj, 126, 2594 

\bibitem[2003]{richards03} Richards, G.~T.~et 
al.\ 2003, \aj, 126, 1131

\bibitem[1999]{schmidt99} Schmidt, G.~D.~\& 
Hines, D.~C.\ 1999, \apj, 512, 125

\bibitem[2003]{schneider03} Schneider, R., Ferrara, A.,
Salvaterra, R., 2003, MNRAS (astro-ph/0307087)

\bibitem[2003]{stern03} Stern, D., Hall, P.~B., 
Barrientos, L.~F., Bunker, A.~J., Elston, R., Ledlow, M.~J., Raines, S.~N., 
\& Willis, J.\ 2003, \apjl, 596, L39

\bibitem[1987]{surdej87} Surdej, J.~\& 
Hutsemekers, D.\ 1987, \aap, 177, 42

\bibitem[2001]{todini01} Todini, P.~\& 
Ferrara, A.\ 2001, \mnras, 325, 726

\bibitem[1997]{umemura97} Umemura, 
M., Fukue, J., \& Mineshige, S.\ 1997, \apjl, 479, L97

\bibitem[2003]{verner03} Verner, E., Bruhweiler, 
F., Verner, D., Johansson, S., \& Gull, T.\ 2003, \apjl, 592, L59

\bibitem[1991]{weymann91} 
Weymann, R.~J., Morris, S.~L., Foltz, C.~B., \& Hewett, P.~C.\ 1991, \apj, 
373, 23

\bibitem[2003]{willott03} Willott, 
C.~J., Rawlings, S., \& Grimes, J.~A.\ 2003, \apj, 598, 909

\end{thebibliography}
\end{document}